\documentclass[preprint,12pt]{elsarticle}
\usepackage[T1]{fontenc}
\usepackage{polski}
\usepackage[slovak, english]{babel}

\usepackage{amssymb}
\usepackage{color}
\usepackage{amsmath}
\usepackage{xspace}
\usepackage{lipsum}
\usepackage{xcolor}
\journal{Chaos, Solitons $\&$ Fractals}

\begin{document}

\begin{frontmatter}



\title{Self-trapping and switching of solitonic pulses in mismatched dual-core highly nonlinear fibers}
\author[label1]{Nguyen Viet Hung}
\ead{hung.nguyenviet@itims.edu.vn}
\affiliation[label1]{organization={International Training Institute for Materials Science (ITIMS), Hanoi University of Science and
Technology (HUST)},%Department and Organization
            addressline={No 1 - Dai Co Viet Str.}, 
            city={Hanoi},
            country={Vietnam}}

\author[label2]{Le Xuan The Tai}

\affiliation[label2]{organization={Faculty of Physics, Warsaw University of Technology},%Department and Organization
            addressline={Koszykowa 75}, 
            city={Warsaw},
            postcode={00-662}, 
            country={Poland}}

\author[label3,label4]{Mattia Longobucco}
\author[label3,label4]{Ryszard Buczyński}
\author[label5,label6]{Ignac Bugár}
\author[label5]{Ignas Astrauskas}
\author[label5]{Audrius Pugžlys} % ž
\author[label5]{Andrius Baltuška}

\affiliation[label3]{organization={Department of Glass, Łukasiewicz Research Network - Institute of Microelectronics  $\&$ Photonics, Aleja},%Department and Organization
            addressline={Lotników 32/46}, 
            city={Warsaw},
            postcode={02-668}, 
            country={Poland}}

\affiliation[label4]{organization={Department of Photonics, Faculty of Physics, University of Warsaw},%Department and Organization
            addressline={Pasteura 5}, 
            city={Warsaw},
            postcode={02-093}, 
            country={Poland}}
        
\affiliation[label5]{organization={Photonics Institute, TU Wien},%Department and Organization
        	addressline={Gu\ss hausstra\ss e 25-29}, 
        	city={Vienna},
        	postcode={1040}, 
        	country={Austria}}

\affiliation[label6]{organization={Department of Chemistry, University of Ss. Cyril and Methodius in Trnava},%Department and Organization
            addressline={Nám. J. Herdu 2}, 
            city={Trnava},
            postcode={917 01}, 
            country={Slovakia}}

\author[label7,label8]{Boris Malomed}

\affiliation[label7]{organization={Department of Physical Electronics, School of Electrical Engineering, Faculty of Engineering, Center
for Light-Matter Interaction, Tel Aviv University},%Department and Organization
            addressline={Tel Aviv 69978}, 
            country={Israel}}

\affiliation[label8]{organization={Instituto de Alta Investigaci\'{o}n, Universidad de Tarapac\'{a}},%Department and Organization
           addressline={Casilla 7D}, 
            city={Arica},
            country={Chile}}

\author[label9]{Marek Trippenbach}

\affiliation[label9]{organization= {Institute of Theoretical Physics, Faculty of Physics, University of Warsaw},%Department and Organization
            addressline={Pasteura 5}, 
            city={Warsaw},
            postcode={02-093}, 
            country={Poland}}

\begin{abstract}
We investigate experimentally and theoretically effects of the inter-core
propagation mismatch on nonlinear switching in dual-core
high-index-contrast soft-glass optical fibers. Incident femtosecond pulses of various energy
are fed into a single (\textquotedblleft straight") core, to identify
transitions between different dynamical regimes, \textit{viz}., inter-core
oscillations, self-trapping in the cross core, and retaining the pulse in
the straight core. The transfer between channels, which has solitonic character, is controlled by the pulse's energy. A
model based on the system of coupled nonlinear Schr\"{o}dinger equations
reveals the effect of the mismatch parameter and pulse duration on the
diagram of the various energy dependent dynamical regimes.
Optimal values of the mismatch and pulse width, which ensure stable
performance of the nonlinear switching, are identified. The theoretical predictions are in agreement with experimental findings.

\end{abstract}




\end{frontmatter}


\section{Introduction}
\label{}
The concept of nonlinear directional couplers based on dual-core fibers
(DCFs) was introduced theoretically in the early 1980s \cite%
{jensen1982,Maier1982}. Since then, considerable efforts were devoted to the
characterization and optimization of the their performance \cite%
{Hui1,Agrawal1,Trillo1}. New perspectives had emerged with the advent of the
photonic-crystal-fiber technology, which offers appropriate conditions for
efficient coherent spectral broadening, especially in the case of anomalous
group-velocity dispersion (GVD) \cite{Herrmann1,Luan1}. Additionally,
theoretical work has predicted that the asymmetry between the cores in the
fiber may be advantageous for the nonlinear switching dynamics in the
soliton propagation regime \cite{He1,Uthayakumar1,Govindarajan1}. The
asymmetry needs to be carefully applied, as an excessively high value of the
inter-core index mismatch destroys the coupling between parallel cores \cite%
{Curilla1}. Therefore, fabrication of photonic crystal fibers (PCFs) with
appropriate properties is a challenging task. Using the PCF production
technology, it is hard to create DCF structures with sufficiently low
asymmetry. As an alternative, a promising candidate for ultrafast pulse
steering was proposed recently, \textit{viz}., a dual-core
high-index-contrast optical fiber made of soft glass \cite{Longobucco1d}.
The simple cladding of such a fiber, surrounding a highly nonlinear core,
ensures, simultaneously, high nonlinearity, tight field localization, and a
low level of dual-core asymmetry at significantly simplified technology \cite%
{Nguyen1}. The applicability of this technology was demonstrated both in a
vicinity of 1700 nm \cite{Longobucco1d} and in the C-band \cite{Nguyen1}.
Parallel to the experimental work, the understanding of the physical
mechanism behind this setting has advanced with the help of extensive
numerical studies \cite{Longobucco1e,Longobucco1a,Malomed1}. The multiple
switching performance, observed under the action of the monotonous increase
of the pulse's energy, has confirmed the role of the soliton self-trapping
in the specialty highly nonlinear fiber \cite{Longobucco1d,Nguyen1}. The
exchangeable self-trapping in both cores is the key mechanism in the studied
dynamical regimes, which could be simulated using a relatively simple model.
The application for real experimental conditions has revealed very good
agreement between the experimental and theoretical results \cite{Nguyen1}.
However, the role of asymmetry of the dual-core structure was not
investigated experimentally or theoretically. In this work we report new
findings, obtained by exciting both cores in the experiment, and introducing
an effective refractive-index mismatch between the cores in the numerical
studies. The reported results provide an essential step forward in the
design of DCFs with an enhanced potential for applications, such as
multiplexers \cite{Miao1} and nonlinear all-optical switches \cite%
{Longobucco1f}.

\section{The theoretical model}

The model is based on the system of linearly coupled nonlinear Schr\"{o}%
dinger equations (NLSEs) \cite{Liu1,Zhao1,Li1}, written for complex
envelopes $A(z,t)$ of electromagnetic waves in the mismatched cores of the
DCF,
\begin{eqnarray}
\partial _{z}A_{1}+\beta _{11}\partial _{t}A_{1}+\frac{i\beta _{21}}{2}%
 \partial _{tt}A_{1}=i\kappa ^{0}_{12} A_{2}-\kappa^{1}_{12} \partial _{t}A_{2} \nonumber \\
 +i\delta A_{1}+i\gamma _{1}|A_{1}|^{2}A_{1}, 
\label{eq:theomodel1}
\end{eqnarray}%
\begin{eqnarray}
\partial _{z}A_{2}+\beta _{12}\partial _{t}A_{2}+\frac{i\beta _{22}}{2}%
\partial _{tt}A_{2}=i\kappa ^{0}_{21} A_{2}-\kappa^{1}_{21} \partial _{t}A_{1} \nonumber \\
 -i\delta A_{2}+i\gamma _{2}|A_{2}|^{2}A_{2}. 
\label{eq:theomodel1a}
\end{eqnarray}%
All coefficients were evaluated at central frequency $\omega _{0}$
corresponding to the wavelength $\lambda _{0}=1700$ nm of the excitation
pulses for the specific fiber employed in our experimental study, using a
mode solver from Lumerical. The two fiber cores have a nearly hexagonal
shape, with the $3.1$ $\mathrm{\mu m}$ distance between their centers and
the effective mode area of $1.66$ $\mathrm{\mu m}^{2}$ at $1700$ nm \cite%
{Nguyen1}. The frequency-independent coupling coefficients $\kappa^{0}_{12}$ and $\kappa^{0}_{21}$ are, respectively:
\begin{eqnarray}
\kappa^{0}_{12} &=&\frac{2\pi^{2}}{\lambda_0^2\beta}\int \int_{-\infty
}^{\infty }(n^{2}-n_{1}^{2})F_{1}^{\ast }F_{2}dxdy,  \label{eq:coupling12} \\
\kappa^{0}_{21} &=&\frac{2\pi^{2}}{\lambda_0^2\beta}\int \int_{-\infty
}^{\infty }(n^{2}-n_{2}^{2})F_{2}^{\ast }F_{1}dxdy.
\end{eqnarray}%
where functions $F_{1}(x,y)$ and $F_{2}(x,y)$ are field-distribution
profiles of fundamental modes in each core, subject to the normalized
conditions,
\[
\int \int_{-\infty
}^{\infty }|F_{1}\left( x,y\right)
|^{2}dxdy=\int \int_{-\infty
}^{\infty }|F_{2}\left(
x,y\right) |^{2}dxdy=1,
\]%
$\kappa^{0}_{21}$ and $\kappa^{0}_{12}$ are the first-order expansion of the frequency dependent coupling coefficient $\kappa$ (coupling dispersion). $n_{1}$ and $n_{2}$ are refractive indices of the two cores, and $n(x,y)$ is
the refractive-index profile of the DCF \cite{Agrawal1}. In our case, the
refractive indices of both cores are identical (the PBG08 glass was used as
the core material, with $n_{1,2}=1.9$), while the asymmetry is underlain by
a difference in shapes of the cores. Beyond the core, the refractive index
is uniform, corresponding to the cladding material, \textit{viz}., UV710
glass ($n=1.52$). The asymmetry parameter is
\begin{equation}
\delta =\frac{1}{2}(\beta _{01}-\beta _{02})  \label{delta}
\end{equation}%
where $\beta _{0m}$ are propagation constants at $\lambda_0$ in the
individual channel ($m=1,2$). The nonlinear Kerr coefficients are:
\begin{equation}
\gamma _{m}=\frac{2\pi\widetilde{n}_{2}}{\lambda_0}\int \int_{-\infty
}^{\infty }|F_{m}(x,y)|^{4}dxdy  \label{eq:Non12}
\end{equation}%
where $\widetilde{n}_{2}=4.3\times 10^{-19}$ $\mathrm{m}^{2}\mathrm{/W}$ is
the nonlinear index of refraction of the PBG08 glass used as the core
material, which is about 20 times higher than in silica.

\section{Rescaling the physical parameters, and an exact solution for the
linearized system}

In the simulations, we used rescaled parameters and noticed that in our fiber differences between the cores in terms of the coupling coefficients are negligible,
therefore: $\kappa^{0,1}_{12} \approx \kappa^{0,1}_{21}=\kappa_{0,1}$. We define dimensionless
parameters for time, distance, and amplitude: $\tau=t\sqrt{%
\kappa_0/|\beta_{21}|}=t/t_0$, $\zeta=z(2\kappa_0/\pi)=z/z_0$, and $\Psi=\sqrt{%
\gamma/\kappa_0}A$, and cast equations (\ref{eq:theomodel1})-(\ref%
{eq:theomodel1a}) in the following form (notice that we have defined unit of
time $t_0$ and unit of length $z_0$, which will later be related to the
pulse duration and propagation length):
\begin{equation}
\label{eq:theomodel3a}
-i\partial_\zeta \Psi_1 =i\epsilon\partial_{T}\Psi_2 -i(\alpha_2-\alpha_1)\partial_T \Psi_1 
+ \frac{1}{2}\partial_{TT} \Psi_1 + \sigma \Psi_1+|\Psi_1|^2\Psi_1+\Psi_2, 
\end{equation}
\begin{equation}
\label{eq:theomodel3b}
-i\partial_\zeta \Psi_2 =
i\epsilon\partial_{T}\Psi_1 + \frac{\alpha}{2}\partial_{TT} \Psi_2 - \sigma \Psi_2 +|\Psi_2|^2\Psi_2 + \Psi_1.
\end{equation}
where $\alpha_1=\beta_{11}/\sqrt{\kappa_0|\beta_{21}|}$, $\alpha_2=\beta_{12}/%
\sqrt{\kappa_0|\beta_{21}|}$, $\alpha=|\beta_{22}|/|\beta_{21}|$, $\epsilon=\kappa_1/\sqrt{\kappa_0|\beta_{21}|}$, and the
mismatched parameter: $\sigma=(\beta_{01}-\beta_{02})/(2\kappa_0)$ and we 
used the retarded time $T=\tau-\alpha_2\zeta$, 
In the experiments, we have achieved the best results at $\lambda_0=1700$
nm, hence all the parameters refer to this wavelength. Below we present a
table with effective values at this wavelength.

\begin{table}[h]
\caption{Optical parameters of the dual-core fiber, which were utilized for
the numerical study of the nonlinear propagation. The parameters,
corresponding to the fiber used in the experiment, were produced with the
help of the mode-solver at the carrier wavelength of $1700$ nm.}
\label{tab:params}
\centering
\bigskip
        \begin{tabular}{|c|c|c|c|c|}
        \hline
        Physical & & & \\
        quantity & 1st core & 2nd core & Units \\
        \hline\hline
        $n_{eff}$ & $1.77766$ & $1.77719$ &  \\
        \hline
        $\beta_0$ & $6.56172\times 10^6$ & $6.55996\times 10^6$ & $1/m$ \\
        \hline
        $\beta_1$ & $6.58061\times 10^{-9}$ & $6.58085\times 10^{-9}$ & $s/m$ \\
        \hline
        $\beta_2$ & $-9.886149 \times 10^{-26}$ & $-9.886149\times 10^{-26}$ & $s^2/m$ \\
        \hline
        $\gamma$ & $0.85338$ & $0.85584$ & $1/(W.m)$ \\
        \hline
        $\kappa_0$ & $1017.8058$ & $1017.8058$ & $1/m$\\
        \hline
        $\kappa_1$ & $ -1.49662\times 10^{-13}$ & $ -1.49662\times 10^{-13}$ & $s/m$\\
        \hline
    \end{tabular}
\end{table}
{Using the values reported in Table \ref{tab:params},} $\alpha_1=656.0245871$, $\alpha_2=656.0485128$, $\epsilon=-0.01492$. Here the last parameter is related to the dispersive character of the coupling coefficient. Even though it seems small in absolute value, it is crucial for pulse propagation dynamics.

Due to the small difference between the GVD and nonlinearity in both cores,
average values were used for the numerical modeling, \textit{viz}., $\beta
_{2}=-9.886149\times 10^{-26}$ $\mathrm{s}^{2}\mathrm{m}^{-1}$ and $\gamma
=0.85461$ $\mathrm{W}^{-1}\mathrm{m}^{-1}$ and $\alpha=1$. It is worth mentioning that the
negative value of $\beta _{2}$ means the anomalous sign of GVD of the fiber
at 1700 nm, hence solitonic propagation may be expected, initiated by the
ultrafast excitation pulse in such a highly nonlinear fiber. 
%In this way, we
%arrive at the final system of propagation equations for our model of the
%double-core DCF with the refractive-index mismatch between the cores:{\color{red} equations need to be correced
%\begin{eqnarray}
%\label{eq:highcore}
%i\partial_\zeta \Psi_1=i(\alpha_2-\alpha_1)\partial_T \Psi_1 \nonumber \\
%&&-\frac{1}{2}\partial_{TT} \Psi_1 -\sigma \Psi_1-|\Psi_1|^2\Psi_1-\Psi_2
%\end{eqnarray}
%\begin{equation}
%\label{eq:lowcore}
%i\partial_\zeta \Psi_2=-\frac{1}{2}\partial_{TT} \Psi_2 +\sigma \Psi_2-|\Psi_2|^2\Psi_2-\Psi_1
%\end{equation}}
%with $\alpha _{2}-\alpha _{1}=0.023921$ (and $\alpha =1$ set in Eq. (\ref%
%{eq:lowcore})). 

The high-index core (the first core) is the one with a high
group velocity, and the low-index core (the second core) with a low group
velocity. 

The units of propagation length and time for our experimental
conditions can be evaluated to be
\begin{equation}
z_{0}=\frac{\pi }{2\kappa }=1.54~\mathrm{mm}
\end{equation}%
\begin{equation}
t_{0}=\sqrt{|\beta _{2}|/\kappa }=9.86~\mathrm{fs}  \label{t0}
\end{equation}

The length of our fiber was about 18 mm, which corresponds to the
dimensionless propagation distance of $18.3$. It is worth mentioning that,
after the completion of full periods of inter-core oscillations in the
linear propagation regime, the initially excited core stays dominant. In
particular, the $18$ mm propagation length, representing about $6$ periods,
maintains this effect, as confirmed experimentally by monitoring the field
distribution in area of the both cores at the output.

The asymmetry parameter $\sigma$ plays an important role in the dynamics of
the pulse propagation in the fiber. Considering the difference between the
optical parameters of the two cores presented in Table \ref{tab:params} is
obvious that the most \textquotedblleft influential" coefficient is the
propagation constant. The group-velocity mismatch, determined by the
frequency derivative of $\beta _{0}$, is more than an order of magnitude
lower, and the GVD mismatch between the cores is completely negligible. For
this reason, the group-velocity difference was fixed, and only $\sigma =
(\beta_{01}-\beta_{02})/(2\kappa)$ was varied in the course of the
simulations, as it represents the dominant effect of the mismatch.
Therefore, in our study, the impact of the asymmetry is investigated by
systematically increasing the value of $\sigma $ from $0$, which represents
the symmetric coupler without any mismatch. The asymmetry parameter is
increased up to the level where the nonlinear switching is still possible,
but with lower sensitivity to small changes of the input energy, in terms of
the output-port-dominance exchange.

We have also examined the effect of the pulse's shape and concluded that the
results are practically the same when \textrm{sech} or Gaussian pulses are
used. The pulse-width effect was examined experimentally in the range
between $110$ and $150$ fs, which is sufficiently broad, taking into
consideration that the soliton order increases linearly with the increasing
width \cite{Agrawal1}. Careful complex amplitude-phase diagnostics was
performed under step-by-step realignment of the setup of the optical
parametric amplification (OPA) source to establish the two above-mentioned
border values: $110$ and $150$ fs. For our simulations, we used the input
Gaussian pulse
\begin{equation}
\Psi (0,\tau )=a\exp \left( -\eta ^{2}\tau ^{2}\right),  \label{Gauss}
\end{equation}
where $a$ is the amplitude of the pulse envelope. From the FWHM definition,
\begin{equation}
\eta \tau =\eta \frac{t_{\mathrm{FWHM}}}{2t_{0}}=\sqrt{\frac{\ln (2)}{2}}%
\approx 0.5887
\end{equation}%
it follows that $\eta =1.1774t_{0}/t_{\mathrm{FWHM}}$, hence the respective
values of the inverse-width parameter in Eq. (\ref{Gauss}) are $\left\{ \eta
(\mathrm{150{\tiny ~}fs});\eta (\mathrm{110{\tiny ~}fs})\right\} =\left\{
(11.609/150);(11.609/110)\right\} =\left\{0.0774;0.1055\right\}$. The energy
of the pulse as a function of $a$ and $\eta $ can be expressed as
\begin{equation}
E=\int_{-\infty }^{+\infty }|A(z,t)|^{2}dt=\frac{\kappa t_{0}}{\gamma }\sqrt{%
\frac{\pi }{2}}\frac{a^{2}}{\eta }=14.739\frac{a^{2}}{\eta }[\mathrm{pJ}].
\label{E}
\end{equation}

The experimental work was carried out with the standard setup presented in
detail in Refs. \cite{Longobucco1d,Longobucco2}. Femtosecond pulses centered
at $1700$ nm were generated in an OPA pumped by the second harmonics of
commercial Yb:KGW laser system (Pharos, Light Conversion) operating at $10$
kHz repetition rate. The OPA allowed the tuning of the pulse wavelength in
the range of $1500-1900$ nm, which is an essential option for studying DCFs
with different levels of the asymmetry. The propagation-constant mismatch
decreases with the increase of the wavelength \cite{Curilla1} therefore the
DCF sample which featured poor switching performance at 1560 nm was studied
in this work, using 1700 nm input pulses. The pulses were guided through a
half-wave plate and polarizer representing a tunable attenuator and through
a second half-wave plate to set the proper pulse polarization. The
in-coupling and out-coupling of the beam were provided by two 40x microscope
objectives mounted on 3D-positioners, securing submicron precision. The
output of the fiber was monitored by an infrared camera imaging the output
facet on its detector surface. Under the single-core excitation, series of camera images were registered by changing the
energy of the excitation pulses in the range of $0.1-1.5$ nJ separately for
the fast and slow core excitation. Additionally, the recordings were
repeated for different pulse widths achieved by tuning the OPA while
simultaneously keeping the central wavelength at $1700$ nm.

It is relevant to mention that the \emph{linearized version} of the system
of Eqs. (\ref{eq:theomodel3a}) and (\ref{eq:theomodel3b}) admits an exact solution
for continuous-wave (CW) states, i.e., ones with constant amplitudes of the
fields:%
\begin{eqnarray}
\left(\Psi _{1}\right) _{\mathrm{CW}} &=&\Psi _{0}\cos \left( K\zeta
\right) e^{i\left(p\zeta -\Omega T\right)} ,  \label{CW1} \\
\left( \Psi _{2}\right) _{\mathrm{CW}} &=&\Psi _{0}\left[ A\cos \left(
K\zeta \right) +iB\sin \left( K\zeta \right) \right] e^{i\left( p\zeta
-\Omega T\right)} .\label{CW2}
\end{eqnarray}%
Here $\Psi _{0}$ is an arbitrary constant amplitude and $\Omega $ is an
arbitrary frequency shift, which defines the family of the CW solutions.
Further, $p$ is the corresponding shift of the propagation constant, $B$ is
the relative amplitude of the waves in the two cores, and $2\pi /K$ is the
period of the power switching between the cores. The latter parameters are
expressed in terms of $\Omega $ as follows:
\begin{eqnarray}
A &=&-\left[ \sigma +\frac{\Omega }{2}\left( \alpha _{1}-\alpha _{2}\right) %
\right]/(1+\epsilon \Omega)  ,  \label{A} \\
B &=&K/(1+\epsilon \Omega) %
 ,  \label{B} \\
p &=&-\frac{1}{2}\Omega ^{2}+\frac{\Omega }{2}\left( \alpha _{1}-\alpha
_{2}\right) ,  \label{cp} \\
K^{2} &=&(1 + \epsilon \Omega )^2 + \left[ \sigma +\frac{\Omega }{2}\left( \alpha _{1}-\alpha _{2}\right)
\right] ^{2}.  \label{ck}
\end{eqnarray}

In the absence of the group-velocity mismatch, i.e., $\alpha _{1}-\alpha
_{2}=0$, this solution is tantamount to the previously known one \cite%
{Agrawal1}. The exact CW solution, given by Eqs. (\ref{CW1})-(\ref{ck}), is
a novel finding. Actually, it can be obtained from the one known for $\alpha
_{1}-\alpha _{2}=0$, if the Doppler shift, $\left( \alpha _{1}-\alpha
_{2}\right) \Omega $, is added to the phase-velocity mismatch, $2\sigma $.

The asymmetry of the solution is characterized by the ratio of the
amplitudes:%
\begin{equation}
\frac{\max \left( \left\vert \Psi _{2}(\zeta )\right\vert \right) }{\max
\left( \left\vert \Psi _{1}(\zeta )\right\vert \right) }=\sqrt{B^{2}+K^{2}}%
\equiv \sqrt{1+2\left[ \sigma +\frac{\Omega }{2}\left( \alpha _{1}-\alpha
_{2}\right) \right] ^{2}},  \label{max}
\end{equation}%
where index 2 represents the excited core. At $\sigma=0.3$ coupling efficiency is still high, as the ration of their amplitudes is equal to $1.086$. Note that the asymmetry is
cancelled at a specially chosen value of the frequency shift,%}
\begin{equation}
\Omega _{0}=-2\sigma /\left( \alpha _{1}-\alpha _{2}\right) .
\label{cancelling}
\end{equation}

Lastly, in the absence of the group-velocity mismatch, $\alpha _{1}-\alpha
_{2}=0$, the simplified form of the exact solution admits a more
sophisticated exact solution. It is a two-component chirped Gaussian pulse, localized (and, in the general case, moving) along the temporal coordinate, with two components periodically oscillating between the cores. These solutions also contain an arbitrary frequency shift $\Omega$, cf. Eqs. (\ref{CW1}) and (\ref{CW2}):%

\begin{eqnarray}
\left( \Psi _{1}\right) _{\mathrm{Gauss}} &=&\Phi (\zeta )\cos \left( \sqrt{%
1+\sigma ^{2}}\zeta \right) \nonumber \\
&&\times \exp \left( -\frac{1}{2}\varphi (z)\left(
T+\Omega \zeta \right) ^{2}-i\Omega T- \frac{i}{2}\Omega ^{2}\zeta \right) ,
\label{Psi1} \\
\left( \Psi _{2}\right) _{\mathrm{Gauss}} &=&\Phi (\zeta )\left[ -\sigma
\cos \left( \sqrt{1+\sigma ^{2}}\zeta \right) +i\sqrt{1+\sigma ^{2}}\sin
\left( \sqrt{1+\sigma ^{2}}\zeta \right) \right]   \nonumber \\
&&\times \exp \left( -\frac{1}{2}\varphi (z)\left( T+\Omega \zeta \right)
^{2}-i\Omega T-\frac{i}{2}\Omega ^{2}\zeta \right) ,  \label{Psi2}
\end{eqnarray}%

where $\varphi (z)$ and $\Phi (z)$ are the following complex functions:%
\begin{eqnarray}
\varphi (z) &=&\frac{1}{W^{2}+i\zeta },  \label{a1} \\
\Phi (z) &=&\frac{W\Psi _{0}}{\sqrt{W^{2}+i\zeta }},  \label{b1}
\end{eqnarray}%
with an arbitrary parameter $W$ which determines the width of the Gaussian.
The oscillation period between the cores is determined by the $\zeta $%
-dependence of the energies of components (\ref{Psi1}) and (\ref{Psi2}), cf.
Fig. \ref{fig:propagation1} below:% {\color{red}
\begin{equation}
\left\{
\begin{array}{c}
E_{1}(\zeta ) \\
E_{2}(\zeta )%
\end{array}%
\right\} =\sqrt{\pi }\left\vert \Psi _{0}\right\vert ^{2}W\left\{
\begin{array}{c}
\sin ^{2}\left( \sqrt{1+\sigma ^{2}}\zeta \right) \\
\sigma ^{2}+\cos ^{2}\left( \sqrt{1+\sigma ^{2}}\zeta \right)%
\end{array}%
\right\} .  \label{E1E2}
\end{equation}%}
Naturally, the total energy, $E_{1}(\zeta )+E_{2}(\zeta )$, stays constant
in the course of the oscillations between the cores.

%\begin{figure}[ht]
%\centering
%\includegraphics[width=1.0\textwidth]{Figure6.png}
%\caption{Simulation results corresponding to the different experimental
%conditions presented as cases (a - top diagram) and (b - bottom diagram) in
%Fig. \protect\ref{fig:experiment1}.}
%\label{fig:experiment2}
%\end{figure}

As concerns the frequency shift, it is related to the experimentally
controllable shift $\Delta \lambda $ of the carrier wavelength, $\lambda_0$.
In physical units, the relation is%
\begin{equation}
\Omega _{\mathrm{phys-units}}\approx -\frac{2\pi c_{0}}{n\lambda_0^{2}}%
\Delta \lambda ,  \label{phys units}
\end{equation}%
where $c_{0}$ is the light speed in vacuum, and $n$ is the refractive index.
In the scaled form adopted above, the value is%
\[
\Omega =t_{0}\Omega _{\mathrm{phys-units}},
\]%
where $t_{0}$ is the time unit defined in Eq. (\ref{t0}). Such simple
analytical approach allows to evaluate the excitation wavelength effect on
the coupling efficiency. In correspondence to our previous works \cite%
{Curilla1} we observed reduced coupling efficiency with decreasing
excitation wavelength, therefore we performed the experimental study at $1700
$ nm instead of $1560$ nm. Using equation (\ref{max}) the same effect is
predicted. Taking into consideration the positive value of $\Delta\lambda$
tuning the wavelength from $1700$ to $1560$ nm and the negative value of
group velocity mismatch $\alpha_1-\alpha_2$, the shorter wavelength
excitation causes reduced coupling efficiency.

\section{Numerical results for nonlinear propagation}

In numerical simulations, we used pulses with two different widths at the
FWHM level, $150$ and $110$ fs to match the experimental data. In the case
of the Gaussian pulse, the corresponding values of the inverse pulse width,
defined above, were $\eta _{1}=0.077$ and $\eta _{2}=0.11$, respectively.
Additionally, we investigated the effect of the mismatch in the effective
refractive index to find optimal conditions for controllable switching
performance. The propagation distance in the simulations was approximately $%
25$ mm, and since the numerically calculated coupling length was 1.54 mm, it
corresponds to about $8$ periods of inter-core-coupling oscillations. It
provides a possibility for analysis of the nonlinear dual-core propagation
even beyond the experimentally studied 18 mm length, and puts the findings
into a broader context. The simulation parameters (see Table \ref{tab:params}%
) were selected from the mode solver analysis of the actual fiber structure.
The use of these parameters allows direct comparison of the numerical
results and experimental observations. In the nonlinear simulations, we considered
both high-index- and low-index-core excitations, resulting in qualitative
agreement with the experimental results in terms of the dependence of the
propagation picture on the input energy. However, the considered simple
model, which takes into account neither the linear dissipative effects
(absorption, Rayleigh scattering), nor nonlinear ones (the stimulated Raman
scattering and the generation of dispersive waves), cannot predict precise values of the
switching energies. Therefore, presenting the numerical results, we refer to
values of the pulse's amplitude, properly comparing the predictions with the
experimental findings.

\begin{figure}[h]
	\centering
	\includegraphics[width=0.7\textwidth]{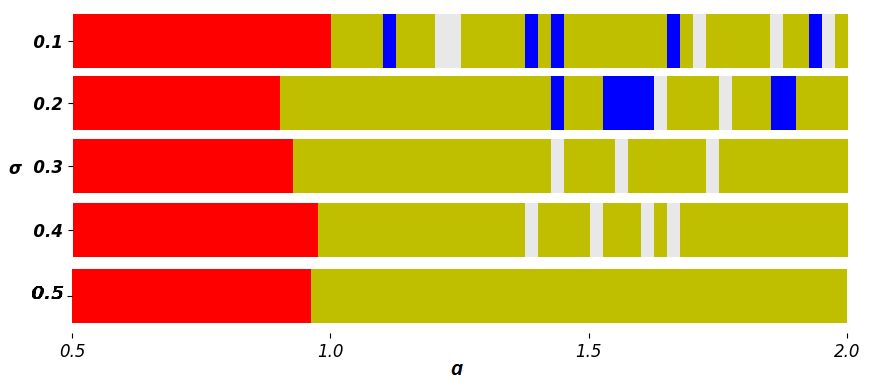}
	\caption{The pulse amplitude dependence of the dynamical propagation regime of the $150$ fs Gaussian pulse in the case of the excitation of the low-index core, for different values of the propagation-constant mismatch, $%
		\protect\sigma $. The red color designates oscillatory behavior, when the final state depends on the actual length of the fiber. Blue means that, after a few initial oscillations, the pulse self-traps mostly in the excited
		(straight) channel; and yellow means the eventual self-trapping in the initially empty (cross) channel. White  stripes were used to mark regions of low contrast, when signal in both channels is comparable, with small oscillations along propagation direction.}
	\label{fig:150fslow}
\end{figure}

\begin{figure}[h]
	\centering
	\includegraphics[width=0.7\textwidth]{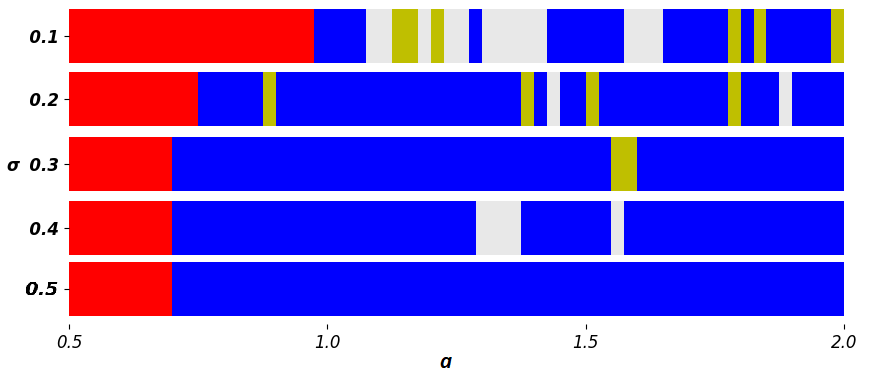}
	\caption{The pulse amplitude dependence of the dynamical propagation regime
		of the $150$ fs Gaussian pulse in the case of the excitation of the
		high-index core, for different values of the propagation-constant mismatch, $%
		\protect\sigma $. The meaning of the color code is the same as in Fig.
		\protect\ref{fig:150fslow}.}
	\label{fig:150fshigh}
\end{figure}

Preliminary experimental observations imply that introducing core asymmetry
may lead to more stable and controllable switching performance
(self-trapping of the pulse in the straight, initially populated or the
opposite, initially empty channel, depending on the initial pulse amplitude)
\cite{Longobucco2}. To put it in quantitative terms, in our simple model we
varied the asymmetry parameter $\sigma$ from $0.1$ to $0.5$ for the $150$ fs
pulse and classified outcomes of the dynamics according to the dependence on
the input pulse amplitude. Results are summarized in Figs. \ref{fig:150fslow}
and \ref{fig:150fshigh}, which represent maps of the nonlinear dynamical
scenarios in two cases, when the incident pulse excites either the low- or
high-index core. The red color designates oscillatory behavior, when the
final state depends on the actual length of the fiber. Blue means that,
after a few initial oscillations, the pulse self-traps mostly in the excited
(straight) channel; and yellow means the eventual self-trapping in the
initially empty (cross) channel. We also marked (with white stripes) cases where we observed low contrast oscillatory behavior as a function of propagation distance. In the case of launching the pulse into the
low-index core, at relatively low mismatch values ($\sigma<0.3$) we observe
several alternations of the pulse trapping between both channels with
increasing amplitude. At some amplitudes of input pulses, after a transient distance the signal becomes almost equally redistributed between two channels, performing low contrast oscillations (white regions). This outcome seems too fragile for the system to be
used as an all-optical switch. However, at $\sigma=0.3$ they are not very frequent and the self-trapping
takes place in the initially empty (cross) channel in a broad range of pulse
amplitude. Such a behavior is quite natural, in view of the propensity of
light to stay in a medium with higher refractive index. This outcome
persists up to the highest analyzed amplitude of the input, i.e. $a= 2.0$, with some exceptions in narrow amplitude regions (white stripes), where equalized energies were predicted comparing the two channels.

In the case of higher mismatch, i.e. $\sigma = 0.4$, similar behavior is predicted, with some equalized dual-core energy distribution situations, but without retaining effect in the excited core. Thus, the low-index core excitation with $150$ fs pulse width is not optimal for nonlinear switching performance. In the case of low asymmetry level ($\sigma\le 0.2$) the system is unstable: there are several transitions between the excited and cross core self-trapping state with increasing pulse amplitude. On the other hand, the higher asymmetry levels ($\sigma>0.2$) does not express self-trapping in the excited core; therefore, it does not support the effective nonlinear switching performance.

If the high-index core is initially excited, we again observe, at first, oscillations-straight (excited) channel trapping transition in the region of
low energy. When the energy is higher, self-trapping occurs also in the
empty (cross) channel. Such switching behavior to the cross-channel takes
place in some narrow intervals of values of $a$ (e.g., around $1.6$ for the moderate asymmetry, $\sigma=0.3$, which is shown in Fig. \ref%
{fig:150fshigh}). Additionally, the trapping threshold decreases when the
asymmetry increases. The reason for the latter effect is that the initial
asymmetry of the fiber strengthened the trend to the self-trapping in the
high-index core. The higher the initial asymmetry, the lower pulse energy is
sufficient to induce additional asymmetry (discrete self focusing in terms
of the channels) for establishing the self-trapping process. The yellow-colored
areas disappear above $\sigma = 0.3$: only equalized dual-core energy effect is predicted (white stripes) in some narrow amplitude intervals. The reason of this behavior in the case of highest asymmetry level is that the initial asymmetry already prevents the self-trapping in the cross-channel. Therefore, we conclude that
$0.3$ is the optimal mismatch value for switching in the case of 150 fs pulse width and high index core excitation, with clear self-trapping
effect also in the non-excited channel. Thus we have a
robust possibility to control the release of the pulse from a particular
output port, regardless of whether the high or low index core is excited.

Analyzing the numerical results, we have concluded that the optimal value of
the asymmetry parameter is $0.3$ because for higher values of $\sigma$ the self-trapping in the originally non-excited core is not more predicted. The
switching dynamics is different when we excite the low- or high-index-core,
with the cross or straight core self-trapping dominance occurring,
respectively, in the former and latter cases. Furthermore, in the case when
the fiber length in the experimental realization is equal to a multiple of
the inter-core-oscillation period, a different peculiarity is observed in
the transition between the inter-core oscillations and self trapping in the
high-index core. As concerns the dominance of the output core, it is
preserved in the case of the excitation of the high-index core, and, on the
contrary, it is exchanged in the case of the low-index core excitation. In
addition to that, the self-trapping may be switched between the two channels
in narrow intervals of the initial amplitude, as may be concluded from Figs. %
\ref{fig:150fslow} and \ref{fig:150fshigh} at low or moderate levels of the
phase-velocity mismatch. The overall dynamics seems more stable in
comparison to that observed in symmetric or weakly-asymmetric DCF studied
before \cite{Nguyen1}, where the diagram of dynamical regimes was more
intricate, exhibiting stronger sensitivity to small variations both of the
amplitude and pulse width.

\section{Detailed comparison with experimental observations}

Here, we aim to compare predictions of the above theoretical model with the
experimental observations made in a nonlinear DCF, with the structure
expressing optical parameters presented in Table \ref{tab:params}, at
wavelength $1700$ nm. Numerical simulations were performed with parameters
matched to the experimental setup, including the wavelength, shape and
duration of the incident pulse.

\subsection{The core selection effect}

\begin{figure}[h]
	\centering
	\includegraphics[width=0.65\textwidth]{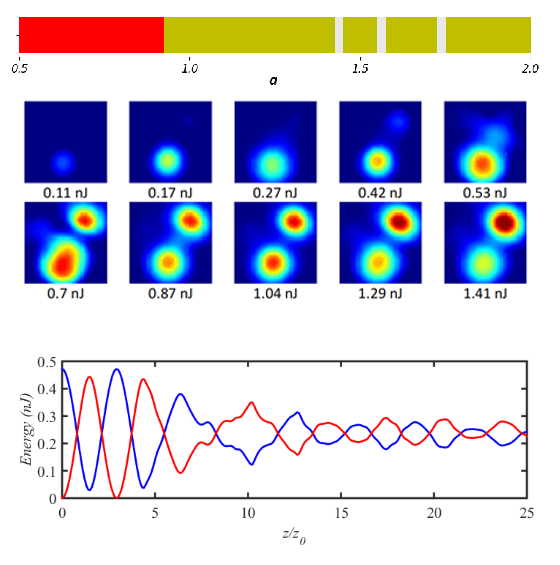}
	\caption{The comparison between the simulations diagram and experimental
		registration of the energy dependent output dual-core field distribution for
		the case of $150$ fs Gaussian pulse launched into the low-index core. Bottom figure shows typical dynamics at the low contrast regions.}
	\label{fig:compare1}
\end{figure}

\begin{figure}[h]
	\centering
	\includegraphics[width=0.7\textwidth]{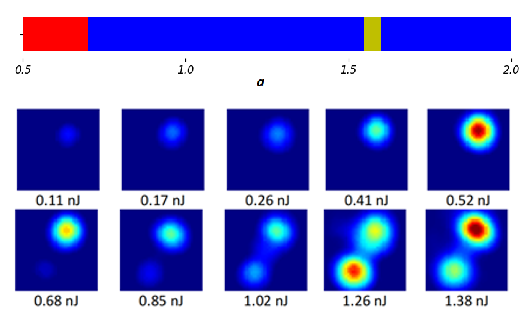}
	\caption{The comparison between the simulations diagram and experimental
		registration of the energy dependent output dual-core field distribution for
		the case of $150$ fs Gaussian pulse launched into the high-index core.}
	\label{fig:compare2}
\end{figure}

In Fig. \ref{fig:compare1} we present the comparison between the theoretical
model and experimental registration for the case of the low-index core
excitation and the incident Gaussian pulse width $t_{FWHM} =150$ fs (the top
panel of Fig. \ref{fig:compare1}). Camera images demonstrate a single
exchange of the dominant core around the critical value of the pulse energy
of $E = 0.87$ nJ. The simulation results predict the same one-step switching
behavior from the inter-core oscillations to self-trapping in the cross
core, which takes place at the amplitude $a = 0.95$. A narrow region of low contrast was predicted around $a = 1.45$, $1.55$ and $1.75$, and a similar effect is visible from the camera images. The bottom panel of Fig. \ref{fig:compare1} reports the distance-dependent dynamics of the energy distribution in both cores (blue - low-index/bottom core, red - high-index/top core) in the case of $150$ fs pulse width and $a = 1.65$. It reveals that the propagation maintains an oscillatory character over the whole analyzed length, with an exponential decrease of the peak power after each period $z/z_0 = 1.65$. It is an example of disturbing effect of the coupling on the soliton self-trapping mechanism. It takes place when the pulse cannot reach the self-trapping critical peak power, resulting in equalized field distribution between the two channels during the solition self-compression process. As a consequence, the self-trapping process doesn't take place and the propagation maintains its harmonic features along the entire considered length. However, such effect require a certain ratio between the self-compression distance determined by the pulse amplitude and the coupling. Therefore, a slight tuning of the amplitude below or above the $1.65$ level result in clear self-trapping effect. In the case of the
high-index-core excitation, the images of the output fiber facet reveal a
different result, viz., transient switching behavior at higher pulse energy,
i.e. around $1.26$ nJ (Fig. \ref{fig:compare2}). Under further increasing of
the pulse's energy, the same straight-core dominance was observed as in the
linear propagation regime. The simulations predict similar outcome with the
transient cross-core dominance effect around the amplitude level of $1.65$.

The transition between the oscillatory and straight-core self-trapping,
predicted by the simulations at the $0.72$ level, is not observable
experimentally because it does not change the core dominance in the output,
due to choice of the fiber's length corresponding to an integer number of
oscillation periods. Thus, the optimal level of the mismatch parameter
identified in the numerical study ($\sigma=0.3$) predicts similar switching performance as the experimental observation. It is a signature of the same optimal
asymmetry level established in the experiment. It was ensured by tuning the
wavelength of the excitation pulses to secure robust switching
performance. Indeed, exciting the same DCF by C-band pulses
resulted in poor switching performance \cite{Longobuccooft}. However, the $1700$ nm excitation improved it significantly; therefore, we set this wavelength also in our
numerical analysis.

\subsection{The pulse-width effect}

\begin{figure}[h]
	\centering
	\includegraphics[width=0.7\textwidth]{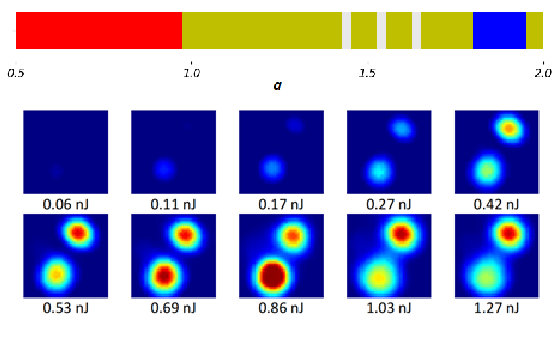}
	\caption{The comparison between the simulations diagram and experimental
		registration of the energy dependent output dual-core field distribution for
		the case of $110$ fs Gaussian pulse launched into the low-index core.}
	\label{fig:experiment1}
\end{figure}

\begin{figure}[h]
	\centering
	\includegraphics[width=0.7\textwidth]{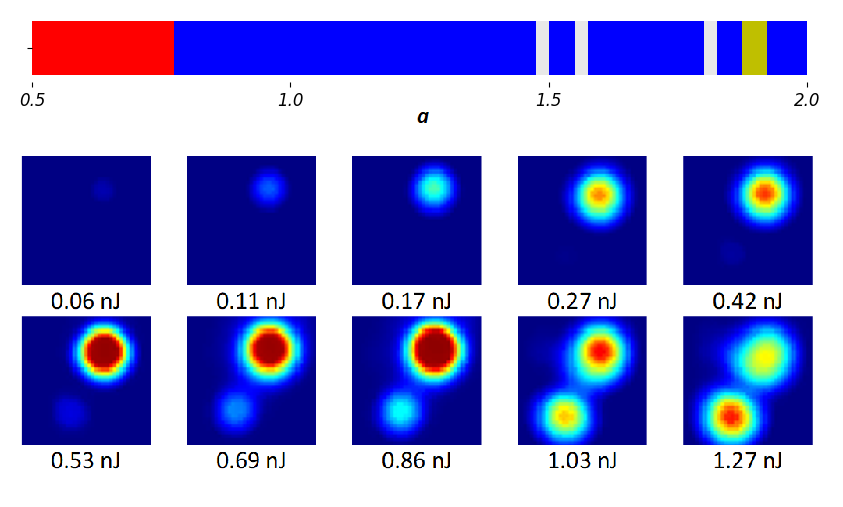}
	\caption{The comparison between the simulations diagram and experimental
		registration of the energy dependent output dual-core field distribution for
		the case of $110$ fs Gaussian pulse launched into the high-index core.}
	\label{fig:experiment2}
\end{figure}

Figure \ref{fig:experiment1} presents the case of the low-index core excitation by pulses with $110$ fs pulse width. It shows a more	sophisticated dynamics than the one in case of $150$ fs pulse width presented in Fig.\ref{fig:150fshigh}. It expresses three transitions between the output
straight/cross core dominance at pulse energies $0.42$ nJ, $0.69$ nJ, and $%
1.03$ nJ, considering both the simulation outcomes (top panel) and the experimental results (bottom panel). The numerical results predict transitions around amplitude levels $0.95$, $1.8$ and $1.95$, which resembles the experimental observations with two dominance exchanges. The images in Figure \ref{fig:experiment2} show the corresponding situation when the high-index core is excited with $110$ fs pulse (bottom panel) and the
predictions of the theoretical model for the same conditions at which the
experiments were performed (top panel). The simulations exhibit three transitions: inter-core oscillations to self-trapping in the straight core at amplitude $0.75$, then some low contrast lines transition to the straight core self-trapping state around level $1.85$, followed by the inverse transition. The latter was not observed in the experiments due to limitations imposed on the pulse input energy, which should be kept below $1.5$ nJ in order not to damage the input facet of the fiber. Furthermore, the oscillations to straight core self-trapping transition does not cause any exchange of the dominant core, as in the case of the results obtained for $150$ fs pulse width (Fig.\ref{fig:150fslow}). Therefore, from the camera images, one can see that only one straight/cross core transition is observed at $1.27$ nJ, in correspondence to the numerical outcomes. between the inter-core oscillations and
self-trapping in the cross core occurs, at the amplitude $0.95$. In
contrast, considering both the low and high-index core excitation cases, the $110$ fs pulse width causes more complex changes in the pulse energy-dependent dual-core propagation dynamics. The reason of such character in the case of the shorter pulse is the linear decrease of the soliton order $N$ with decreasing pulse width, according to equation 

\begin{equation}
	N^2=\frac{\gamma P_0T_0^2}{|\beta_2|},
	\label{eq:n}
\end{equation}

where $P_0$ is the input pulse peak power and $T_0$ the pulse width. The lower soliton order in the case of $110$ fs pulse width reduces the disturbance of the pulse during the soliton fission process. Thus, the more preserved single pulse character supports more exchanges between the trapped channels with increasing pulse energy. The soliton self-compression effect,
characterized by the factor $F_c = 4.1N$ \cite{Agrawal2} is also more pronounced in the case of longer pulses. Consequently, the stronger
selective self-focusing (which favors a particular channel) prohibits the
transfer to the straight core at higher pulse amplitudes. Summarizing this sub-chapter, the $110$ fs pulse width seems to be more advantageous because it enables high switching contrast between the channels based on self-trapping taking place in both of them. It is governed just by a slight change of the pulse amplitude and it is predicted in the case of both high- and low-index core excitation. The experimental observations confirmed these findings comparing $110$ fs vs. $150$ fs pulse excitation and considering both cores excitation.

\begin{figure}[h]
	\centering
	\includegraphics[width=0.7\textwidth]{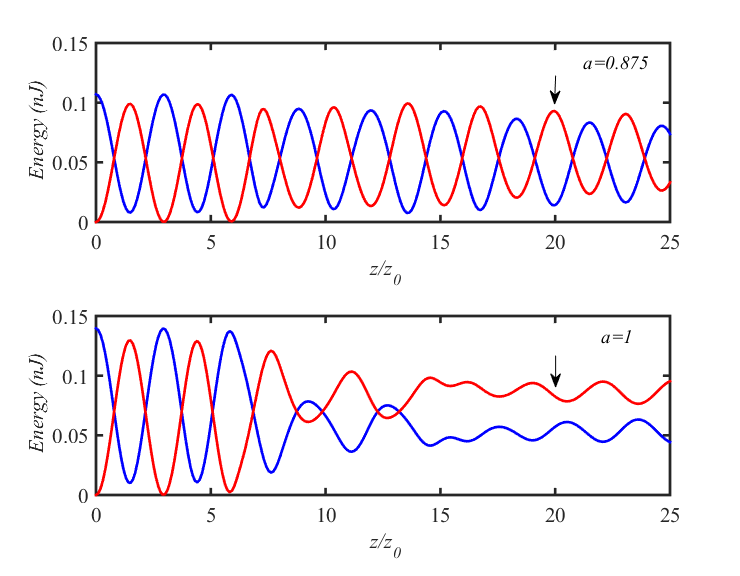}
	\caption{The dependence of the integral field energy on the propagation
		distance in both cores, as produced by the simulations. It shows the
		transition of oscillations to the cross core self-trapping for the input
		pulse amplitudes $0.875$ (the upper panel) and $1.0$ (the lower panel).
		Excitation pulses with width of $110$ fs and Gaussian shape were launched
		into the low-index core of the fiber with asymmetry parameter $\protect\sigma%
		=0.3$. The black arrow marks the length of the fiber in the experiment.}
	\label{fig:propagation1}
\end{figure}

\begin{figure}[h]
	\centering
	\includegraphics[width=0.7\textwidth]{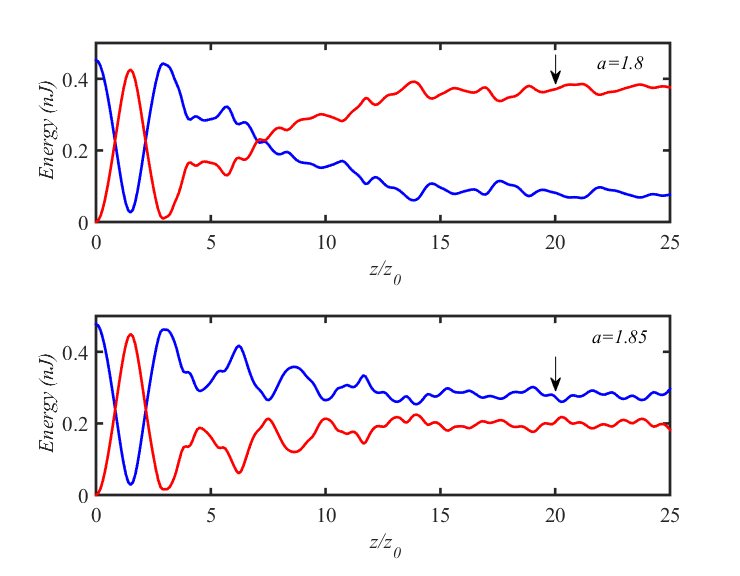}
	\caption{The dependence of the integral field energy on the propagation
		distance in both cores, as produced by the simulations. It shows the
		transition of the trapping from the cross core to the straight one for input
		pulse amplitudes $1.8$ (the upper panel) and $1.85$ (the lower panel).
		Excitation pulses with width of $110$ fs and Gaussian shape were launched
		into the low-index core of the fiber with asymmetry parameter $\protect\sigma%
		=0.3$.}
	\label{fig:propagation2}
\end{figure}

\begin{figure}[h]
	\centering
	\includegraphics[width=0.7\textwidth]{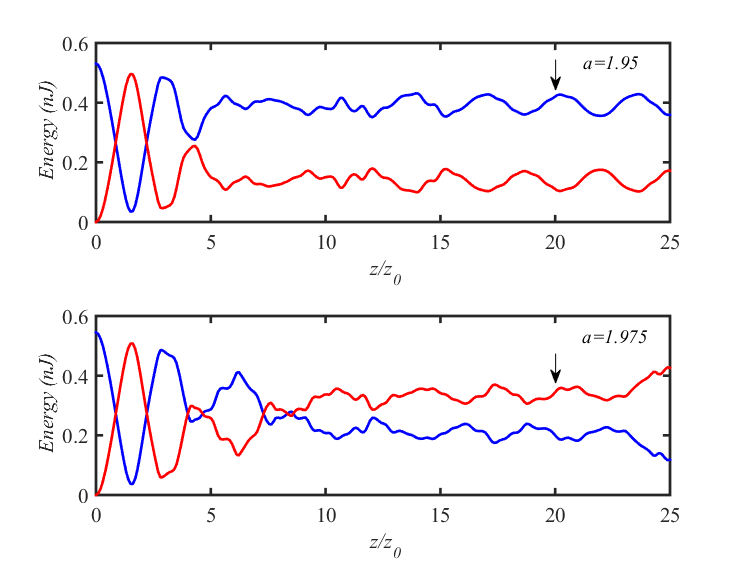}
	\caption{The dependence of the integral field energy on the propagation
		distance in both cores, as produced by the simulations. It shows the
		transition of the trapping from the straight core to the cross one for input
		pulse amplitudes $1.95$ (the upper panel) and $1.975$ (the lower panel).
		Excitation pulses with width of $110$ fs and Gaussian shape were launched
		into the low-index core of the fiber with asymmetry parameter $\protect\sigma%
		=0.3$.}
	\label{fig:propagation3}
\end{figure}

Finally, in Figs.\ref{fig:propagation1}-\ref{fig:propagation3} we
illustrate the propagation-distance-dependent distribution in the both cores
in the case of $110$ fs pulse width, which signals the onset of the same
sequence of three transitions as observed experimentally (bottom panel of Fig. %
\ref{fig:experiment1}). The top panel of Fig. \ref{fig:propagation1} reveals
that, at the low amplitude level, when nonlinear effects are small, the propagation features oscillatory
character in the whole studied propagation range \cite{Nguyen1}. In the propagation evolution graphs
with higher pulse amplitudes, the harmonic behavior terminates after few
initial oscillations due to the soliton self-compression and the subsequent
self-trapping process \cite{Longobucco1a}. According to our simulations,
when the self-trapping commences, the core dominance is preserved in the
whole subsequent range of the studied propagation lengths, including also
the value which corresponds to the fiber length in the experiment. Another
important aspect of the dual-core field-evolution plots is that they express
nearly $100\%$ transfer of the pulse's energy between the cores. It
originates from the low level of the propagation constant mismatch parameter, $0.3$, which,
according to Eq. (\ref{max}) describing various linear
propagation approaches causes only a slight modification of the effective
coupling constant, hence the coupling period remains similar to that for the
zero mismatch. Black arrows indicate the observation point, which
corresponds to the fiber length used in the experiment. All three
transitions presented subsequently in Figs. \ref{fig:propagation1}-\ref%
{fig:propagation3} exhibit a clear exchange of the dominant cores following
a slight increase of the pulse amplitude between the top and bottom panel.
Accordingly, all of them have been identified experimentally by the camera
monitoring the output fiber facet, and the corresponding pairs of camera
images ($0.27$ - $0.42$ nJ, $0.53$ - $0.69$ nJ, $0.86$ - $1.03$ nJ) exhibit
convincing switching contrasts. Thus, the experimental observations have
confirmed the predictions of the numerical simulations, i.e. the
three-transition character of the energy dependence for the $110$ fs pulse
width, in the case of low-index core excitation.

\section{Conclusion}
In conclusion, by means of systematic numerical simulations supported also
by experimental findings, we have studied the switching dynamics in
asymmetric nonlinear DCFs (dual-core fibers), with the propagation-constant
mismatch between the guiding channels. We used a relatively simple numerical
model taking into account all essential parameters of the fiber and input
pulses, tailored to match the experimental conditions. An important control
parameter is the inter-core difference of the effective refractive index
determining the propagation-constant mismatch, which can be determined from
the cross-section image of the DCF structure. However, such estimation
bears high level of uncertainty, because it is defined by the overlap
integrals of the field distribution in the two cores \cite{Agrawal1}.
Therefore, its value is sensitive to fluctuations of the fiber
microstructure along the propagation direction, even at the nanometer scale.
On the other hand, in numerical simulations of the nonlinear propagation we
introduce average mismatch parameter $\sigma$ and determine its optimum
value to be $\sigma=0.3$. The simulations confirm experimental observations
in terms of energy-dependent pulse dynamics when changing the excited core
(low- or high-index), and tuning the inverse pulse width $\eta$. Note also that
previously, good agreement was found in the study of the soliton propagation
in the symmetric DCF \cite{Nguyen1}. Here the theoretical and experimental
findings are summarized in back-to-back maps and camera images showing the
optical field at the end of the fiber as functions of amplitude and energy
of the excitation pulse, respectively. The comparison demonstrates only qualitative
agreement between experimental findings and theoretical results, which were
produced by the simplified model. As mentioned above, this study is similar
to the case of the symmetrical coupler \cite{Nguyen1}, where we could
properly identify individual steps of the dominance exchange between the two
guiding cores. However, in the recent work, we observe much more robust and
controllable switching dynamics, thanks to the effect of higher level of
dual-core asymmetry. The outcomes as concerns the numerical and experimental
approach alike, reveal a single transition between the inter-core
oscillations and self-trapping in the cross core in the case of the
low-index-core excitation and pulse width of $150$ fs. In the case of the
high-index-core excitation by pulses of the same width both simulations and
experiments reveal two transitions (Fig.\ref{fig:compare2}), and the
low-index-core excitation by $110$ fs pulses reveal three transitions (Figs.\ref{fig:experiment1}-\ref{fig:experiment2}),
following the increase of the pulse's amplitude and energy. The predictions
of the numerical model are correct, with the fiber length of $18$ mm,
representing about $6$ full dual-core oscillation periods.

These outcomes reveal an advantage of slight dual-core asymmetry, as
compared to the totally symmetric DCF structures, in terms of more robust
switching dynamics, which is resistant to small fluctuations of the input
pulse's energy. It is a significant progress in the case of applications of
all-optical binary operation, where the switching is controlled establishing
two different input amplitude levels.

It is relevant to stress, that the identified optimal asymmetry level still
represents just a slight perturbation on top of the propagation in the
symmetric coupler, as observed in Figs. \ref{fig:propagation1}-\ref%
{fig:propagation3}. Therefore, the experimental study required a low level
of the dual-core asymmetry, which became available only recently, provided
by new technology based on use of the high-index-contrast soft glass. Hence,
the results offer an essential added value to the further designing of novel
all-solid DCFs intended for all-optical switching purposes. We have also
improved the numerical model: besides considering the propagation constant and group velocity mismatch between the guiding channels, the coupling coefficient dispersion was added as well. It turned out to be an efficient tool for
analyzing the pulse-width effect on the switching in the femtosecond
soliton-propagation regime. Further benefit of this work is the analytic solution of the mismatched coupler in the linear propagation regime. It confirms that: a) at $\sigma = 0.3$ we observe only just a slight perturbation to the linear propagation in terms of coupling efficiency and period, b) with increasing wavelength the asymmetry is decreasing. In our numerical studies we went beyond the analytical approach, by adding the nonlinear and dispersion terms, however neglecting the higher-order nonlinear effects. Therefore the dynamics in our simulations, which is in agreement with the experimental observations at low energies, still preserve the features of the analytical solution and predicts the more robust nonlinear switching behavior at slight enhancement of the asymmetry.

Conditions concerning the pulse's energy and width, which are required of
our approach, are in agreement with parameters of the commercial ultrafast
oscillators. Thus, using our numerical tool, we can predict the proper level
of the propagation-constant mismatch, taking into consideration the wavelength,
pulse's width and top energy provided by such sources, in order to establish
robust, high-contrast switching of the pulses. For example, it is possible
to propose a mismatch parameter at $1560$ nm, which will result in similar
advantageous switching behavior as we obtained in this work performed at $1700$
nm. Combining our novel specially designed dual-core fibers with such
relatively cheap lasers opens the way for development of the synchronization
of the ultrafast oscillators, sampling of data-carrying signals, or their
redirection with the transfer rate exceeding 1 Tb/s. An additional asset of
the model is its applicability to the design of fiber couplers with the
appropriate mismatch, considering parameters of commercially available
ultrafast laser sources.

\section*{Acknowledgment}
This work was supported by the Polish National Science Center under projects
with No. 2020/02/Y/ST7/00136 (R.B.), 2016/22/M/ST2/00261 (M.T., R.B., M.L.)
and 2019/33/N/ST7/03142 (M.L.), by the Vietnam Ministry of Education and
Training (MOET) under Grant Number B2022-BKA-14 (N.V.H.), by the Austrian
Science Fund (FWF) under Grant number I 5453-N (I.B., A.P., A.B.) and by Slovak Scientific Grant Agency through grant No. VEGA 2/0070/21 (I.B.) and, in
part, by the Israel Science Foundation through grant No. 1695/22 (B.A.M). The authors declare no conflicts of interest.



\begin{thebibliography}{00}

\bibitem{jensen1982}
S.~Jensen, \emph{The nonlinear coherent coupler}, IEEE J. Quantum Electron. 18, 1580–1583, 1982.

\bibitem{Maier1982}
A. A. Maier, \emph{Optical transistors and bistable devices utilizing nonlinear transmission of light in systems with
undirectional coupled waves,} Sov. J. quantum Electron. 12, 1490–1494, 1982.

\bibitem{Hui1}
R. Hui, \emph{Chapter 6 - passive optical components,} in \emph{Introduction to Fiber-Optic Communications}, 1st ed, (Elsevier, 2020, pp. 209–297).

\bibitem{Agrawal1}
G. P. Agrawal, \emph{Fiber couplers,} in \emph{Applications of Nonlinear Fiber Optics}, (Academic Press, 2008, pp. 54–99).

\bibitem{Trillo1}
S. Trillo, E. M. Wright, G. I. Stegeman, and S. Wabnitz, \emph{Soliton switching in fiber nonlinear directional couplers,} Opt. Lett. 13, 672, 1988.


\bibitem{Herrmann1}
J. Herrmann, U. Griebner, N. Zhavoronkov, A. Husakou, D. Nickel, J. C. Knight, W. J. Wadsworth, P. S. Russell,
and G. Korn, \emph{Experimental evidence for supercontinuum generation by fission of higher-order solitons in photonic
fibers,} Phys. Rev. Lett. 88, 1739011–1739014, 2002.

\bibitem{Luan1}
F. Luan, A. Yulin, J. C. Knight, and D. V. Skryabin, \emph{Polarization instability of solitons in photonic crystal fibers,} Opt. Express 14, 6550, 2006.


\bibitem{He1}
X. He, K. Xie, and A. Xiang, \emph{Optical solitons switching in asymmetric dual-core nonlinear fiber couplers,}
Optoelectronics Adv. Mater. - Rapid Commun. 4, 3, March 2010 4,3, 284–286, 2010.

\bibitem{Uthayakumar1}
T. Uthayakumar, R. V. J. Raja, K. Nithyanandan, and K. Porsezian, \emph{Designing a class of asymmetric twin core
photonic crystal fibers for switching and multi-frequency generation,} Opt. Fiber Technol. 19, 556–564, 2013.

\bibitem{Govindarajan1}
A. Govindarajan, B. Malomed, A. Mahalingam, and T. Uthayakumar, \emph{Modulational instability in linearly coupled
asymmetric dual-core fibers,} Appl. Sci. 7, 645, 2017.

\bibitem{Curilla1}
L. Curilla, I. Astrauskas, A. Pugzlys, P. Stajanca, D. Pysz, F. Uherek, A. Baltuska, and I. Bugar, \emph{Nonlinear
performance of asymmetric coupler based on dual-core photonic crystal fiber: Towards sub-nanojoule solitonic
ultrafast all-optical switching,} Opt. Fiber Technol. 42, 39–49, 2018.

\bibitem{Longobucco1d}
M. Longobucco, I. Astrauskas, A. Pugžlys, D. Pysz, F. Uherek, A. Baltuška, R. Buczynski, and I. Bugár, \emph{Broadband
self-switching of femtosecond pulses in highly nonlinear high index contrast dual-core fibre,} Opt. Commun. 472,
126043, 2020.

\bibitem{Nguyen1}
V. H. Nguyen, L. X. T. Tai, I. Bugar, M. Longobucco, R. Buczyński, B. A. Malomed, and M. Trippenbach, \emph{Reversible
ultrafast soliton switching in dual-core highly nonlinear optical fibers,} Opt. Lett. 45, 5221, 2020.

\bibitem{Longobucco1e}
M. Longobucco, P. Stajanča, L. Čurilla, R. Buczynski, and I. Bugár, \emph{Applicable ultrafast all-optical switching by
soliton self-trapping in high index contrast dual-core fiber,} Laser Phys. Lett. 17, 025102, 2020.

\bibitem{Longobucco1a}
M. Longobucco, J. Cimek, L. Čurilla, D. Pysz, R. Buczyński, and I. Bugár, \emph{All-optical switching based on soliton
self-trapping in dual-core high-contrast optical fibre,} Opt. Fiber Technol. 51, 48–58, 2019.

\bibitem{Malomed1}
B. A. Malomed, \emph{A variety of dynamical settings in dual-core nonlinear fibers,} in Handbook of Optical Fibers,
(Springer, 2019, pp. 421–474).

\bibitem{Miao1}
W. Miao, F. Yan, and N. Calabretta, \emph{Towards petabit/s all-optical flat data center networks based onwdm optical
cross-connect switches with flow control,} J. Light. Technol. 34, 4066–4075, 2016.

\bibitem{Longobucco1f}
M. Longobucco, I. Astrauskas, A. Pugžlys, D. Pysz, F. Uherek, A. Baltuška, R. Buczynski, and I. Bugár, \emph{High
contrast all-optical dual wavelength switching of femtosecond pulses in soft glass dual-core optical fiber,} J. Light.
Technol. 39, 5111–5117, 2021.

\bibitem{Liu1}
M. Liu and K. S. Chiang, \emph{Propagation of ultrashort pulses in a nonlinear two-core photonic crystal fiber,} Appl.
Phys. B Lasers Opt. 98, 815–820, 2010.

\bibitem{Zhao1}
J. Zhao, Z. Wang, Y. Liu, and B. Liu, \emph{Switchable-multi-wavelength fiber laser based on dual-core all-solid photonic
bandgap fiber,} Front. Optoelectron. China 3, 283–288, 2010.

\bibitem{Li1}
J. H. Li, K. S. Chiang, and K. W. Chow, \emph{Switching of ultrashort pulses in nonlinear high-birefringence two-core
optical fibers,} Opt. Commun. 318, 11–16, 2014.

\bibitem{Longobucco2}
M. Longobucco, I. Astrauskas, A. Pugžlys, N. T. Dang, D. Pysz, F. Uherek, A. Baltuška, R. Buczyński, and I. Bugár,
\emph{Complex study of solitonic ultrafast self-switching in slightly asymmetric dual-core fibers,} Appl. Opt. 60, 10191, 2021.

\bibitem{Agrawal2}
G. P. Agrawal, \emph{Pulse compression,} in \emph{Applications of Nonlinear Fiber Optics}, (Academic Press, 2008, pp. 263–318).

\bibitem{Longobuccooft}
M. Longobucco, J. Cimek, D. Pysz, R. Buczyński, and I. Bugár,
\emph{All-optical switching of ultrafast solitons at 1560 nm in dual-core fibers with high contrast of refractive index,} Opt. Fiber Technol. 63, 102154, 2021.

\end{thebibliography}


\end{document}